\documentclass[%
 reprint,
superscriptaddress,
 amsmath,amssymb,
 aps,
prb,
]{revtex4-1}
\usepackage[english]{babel}
\usepackage{mathtools}
\addto\captionsenglish{}
\usepackage{textcomp}
\usepackage{graphicx}
\graphicspath{{C:/Users/Sudesh/Desktop/NbPArXiv/Figures/}}
\DeclareGraphicsExtensions{.png,.pdf,.tif,.jpeg}

\usepackage{epstopdf}
\AppendGraphicsExtensions{.tif}

\usepackage{dcolumn}
\usepackage{bm}

\begin{document}


\title{Evidence for trivial Berry phase and absence of chiral anomaly in semimetal NbP}

\author{Sudesh}
\email{bsudesh02@gmail.com}
\affiliation{School of Physical Sciences, Jawaharlal Nehru University, New Delhi-110067, India.}
\author{P. Kumar}
\affiliation{School of Physical Sciences, Jawaharlal Nehru University, New Delhi-110067, India.}
\author{P. Neha}
\affiliation{School of Physical Sciences, Jawaharlal Nehru University, New Delhi-110067, India.}
\author{T. Das}
\affiliation{Department of Physics, Indian Institute of Science, Bengalore-560012, India}
\author{A. K. Rastogi}
\affiliation{School of Physical Sciences, Jawaharlal Nehru University, New Delhi-110067, India.}
\author{S. Patnaik}
\email{spatnaik@mail.jnu.ac.in}
\affiliation{School of Physical Sciences, Jawaharlal Nehru University, New Delhi-110067, India.}

\date{\today}

\begin{abstract}
We report a detailed magneto-transport study in single crystals of NbP.   High quality crystals were grown by vapour transport method. An exceptionally large magnetoresistance is confirmed at low temperature (5.4$ \times $10$ ^{4}\%$ at 2.5 K and 6 T) which is non-saturating and is linear at high fields. Models explaining the linear magnetoresistance are discussed and it is argued that in NbP this is linked to charge carrier mobility fluctuations. Negative longitudinal magnetoresistance is not seen, unlike several other Weyl mono-pnictides, suggesting lack of well defined chiral anomaly in NbP. Unambiguous Shubnikov-de-Haas oscillations are observed at low temperatures that are correlated to Berry phases. The Landau fan diagram indicates trivial Berry phase in NbP crystals corresponding to Fermi surface extrema at 30.5 Tesla.  
\begin{description}
\item[PACS numbers]03.65Vf, 72.15Gd, 71.20.Gj
\item[Keywords]Topological phases, Magnetoresistance, Weyl semimetal

\end{description}
\end{abstract}

\pacs{Valid PACS appear here}
\maketitle


\section{\label{sec:level1}Introduction}

The massless solutions of Dirac equations in relativistic particle physics have recently found material basis in topological phases of quantum condensed matter, and this has led to rediscovery of several semimetals with such exotic quantum field theory perspectives \cite{R1, R2, R3}. From the experimental point of view, perhaps the starting point was the Quantum Hall effect where in the role of magnetic field is replaced by strong spin-orbit coupling in topological insulators of Bismuth based chalcogenides \cite{R4,R5}.  The peculiar topology of the band structure of such tetradymites leads to a fully gapped insulating bulk and symmetry-protected conducting surface states \cite{R5, R6, R7}. In the very recent past, similar properties are observed in a new class of topological quantum phase in the Weyl semimetals (WSMs) \cite{R8}; the so called 3D analogues of Graphene \cite{R9}. While WSMs are not gapped in bulk, they have gapless nodes (band-touching points) distributed in three dimensional (3D) momentum (\textbf{k}) space. The linear dispersion at these nodes results in Dirac fermions, but with either time reversal or spatial inversion symmetry breaking, the degenerate Dirac point splits into two Weyl points.  The nodes always appear in pairs with opposite chirality defined by their opposite Chern numbers in the Brillouin zone \cite{R8, R10}. In the electronic band structure, the consequence of this is the presence of Fermi arcs terminating at Weyl points \cite{R11,R12}, that conjures remarkable electromagnetic properties such as anomalous Hall effect \cite{R13, R14},  chiral magnetic effects \cite{R15} and negative longitudinal magnetoresistance \cite{R16}.\par
A major aspect of current debate concerning Weyl semimetals is the precise topological characteristics of the band structure in the Brillouin zone and its experimental manifestations. It has been suggested for long that the Berry curvature in momentum space leads to the topological classification of metallic systems. More precisely, the Weyl points in the band structure are construed analogous to monopoles of Berry curvature and constitute the sinks and sources of Berry flux \cite{R17, R18}. The Berry phase is introduced to account for the cyclotron motion of Weyl fermions in the presence of magnetic field. A zero Berry phase factor implies topologically trivial energy band while a non-zero $ \pi $-Berry phase suggests topologically non-trivial energy band. Experimentally the conclusion on Berry phase is assessed by Shubnikov de-Haas (SdH) oscillations by mapping the Landau-Level (LL) fan diagram (LL index, n vs 1/H). Quite generally, a non-trivial $ \pi $-Berry phase that accounts for additional geometrical phase factor, along a closed trajectory enclosing a Weyl node, is predicted for several WSMs \cite{R17, R18}. However in NbP, the experimental reports are contradictory to one another. Recently, Wang et al. \cite{R19} have reported negative longitudinal magnetoresistance due to chiral anomaly while results of Shekhar et al. \cite{R20} indicate absence of chiral anomaly in NbP. Using quantum oscillation measurements, in this paper, we address these issues and investigate the origin of exceptional magnetoresistance in NbP and its dependence on chiral anomaly and Berry phase factor.\par 

Most investigations of recent past has focused on the study of binary WSMs derived from transition metal monopnictides (NbAs, NbP, TaAs and TaP) \cite{R10, R11, R21, R22, R23, R24}. All these compounds possess non-centrosymmetric tetragonal structure with space group I4$ _{1} $md. The Weyl nodes in these cases are created due to spatial inversion symmetry breaking. Amongst these compounds, NbP shows extremely large magnetoresistance (MR) and ultrahigh mobility \cite{R20}. In essence it combines electronic band structure features of WTe$ _{2} $ and Cd$ _{3} $As$_{2} $-type semimetals, that results in normal quadratic bands from hole pockets and linear Weyl bands from electron pockets. The search of chiral effect and negative longitudinal magnetoresistance in WSMs has collaterally led to the observation of extremely high positive transverse (\textit{H}$ \bot $\textit{I}) magnetoresistance \cite {R20}. \par  
In this paper, we report synthesis and extensive characterization of single crystals of Weyl semimetal NbP through magneto-transport measurements. An extremely large MR is observed at low temperatures which is non-saturating and linear at high fields. This linear MR scales well with the average mobility of charge carriers. In conjunction with Hall data the linear MR is attributed to the mobility fluctuations induced by scattering from low mobility inhomogeneous islands. Further, study of Shubnikov-de Haas oscillations suggests trivial Berry phase in NbP.

\section{\label{sec:level2}Experimental Details}
Millimeter size single crystals of NbP were synthesized using vapour transport technique. In the first step, polycrystalline samples of NbP were synthesized using solid state reaction method. Nb powder (Sigma Aldrich, 99.8\%) and Phosphorus chips (Sigma Aldrich, 99.999\%) were ground together and pressed into pellets using a hydraulic press. The pellets were then vacuum sealed in quartz tubes and sintered at 850\textdegree C for 48 hours. The polycrystalline NbP was then sealed with iodine (13 mg/cm$ ^{3} $) (Sigma Aldrich, ≥ 99.99\%) in a quartz tube of length 12 cm and ID = 12 mm. The tube was put in a tubular furnace which is calibrated to have a temperature gradient of 100\textdegree C  over a distance of 12 cm when set at 950\textdegree C. The samples were sintered for 2 weeks. The obtained crystals were ground and characterized with powder X-ray diffraction at room temperature using a Rigaku X-ray powder diffractometer (Miniflex-600, Cu-K$\alpha$). Energy Dispersive X-ray analysis (EDX) was performed using Bruker AXS microanalyzer. High resolution transmission electron microscopy (HRTEM) measurements were performed using JEOL (JEM-2100F) transmission electron microscope. Magnetoresistivity and Hall measurements were done using a \textit{Cryogenic} cryogen Free Magnet (CFM) system.
\begin{figure}[h]
\centering
\begin{minipage}{0.45\textwidth}
\centering
\includegraphics[width=1.0\textwidth]{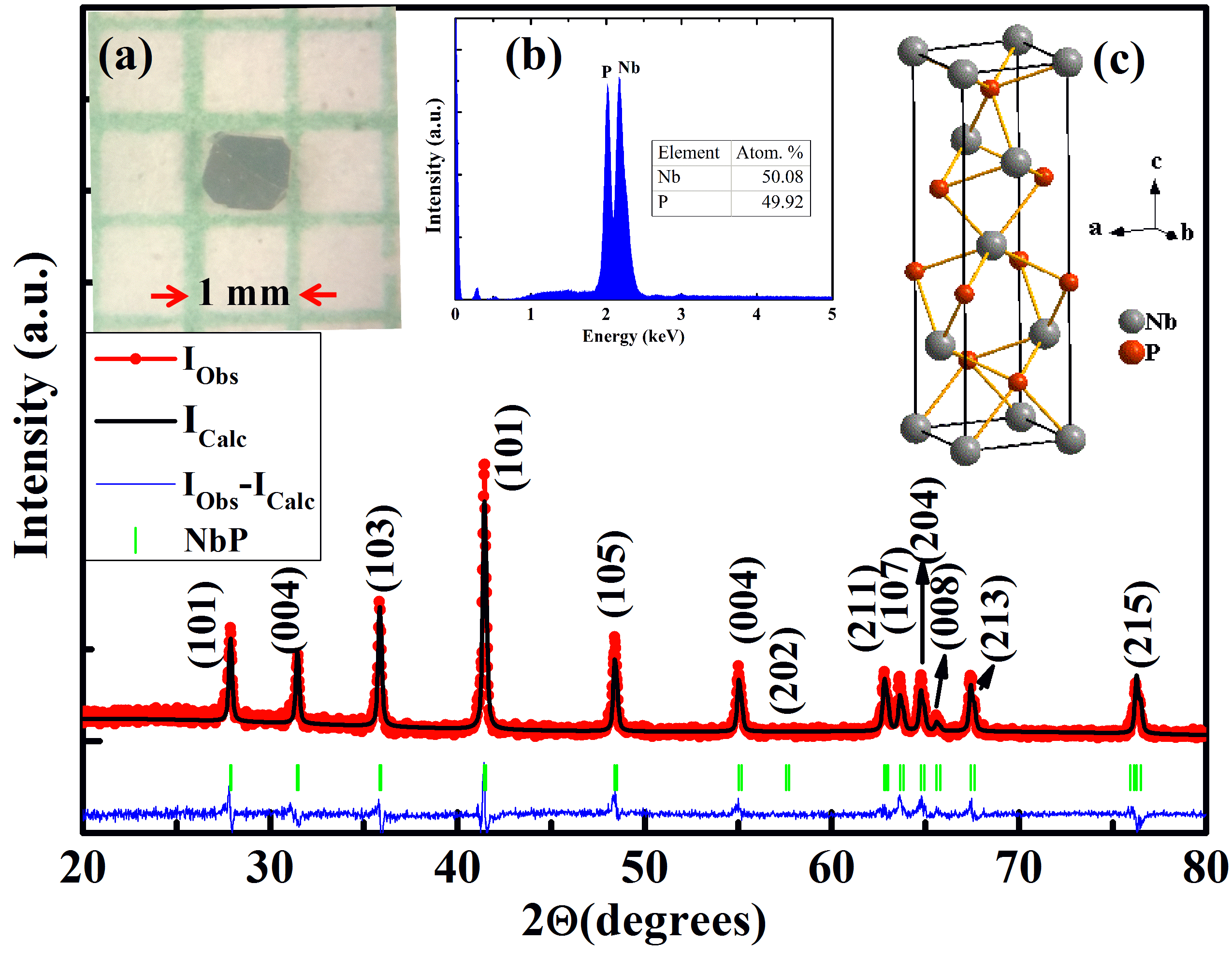}
\caption{Powder X-ray diffraction pattern of NbP single crystals and its rietveld refined data are shown. X-ray diffraction is done after grinding a few single crystals of NbP. Inset (a) shows the image of a typical single crystal sample of NbP at the millimeter scale. Inset (b) shows the EDAX data which confirms stoichiometry of the sample. The schematic of the unit cell of NbP is shown in inset (c).}
\end{minipage}
\end{figure}
\par

\section{\label{sec:level3}Results and Discussion}

\begin{figure}
\centering
\begin{minipage}{0.45\textwidth}
\includegraphics[width=\textwidth]{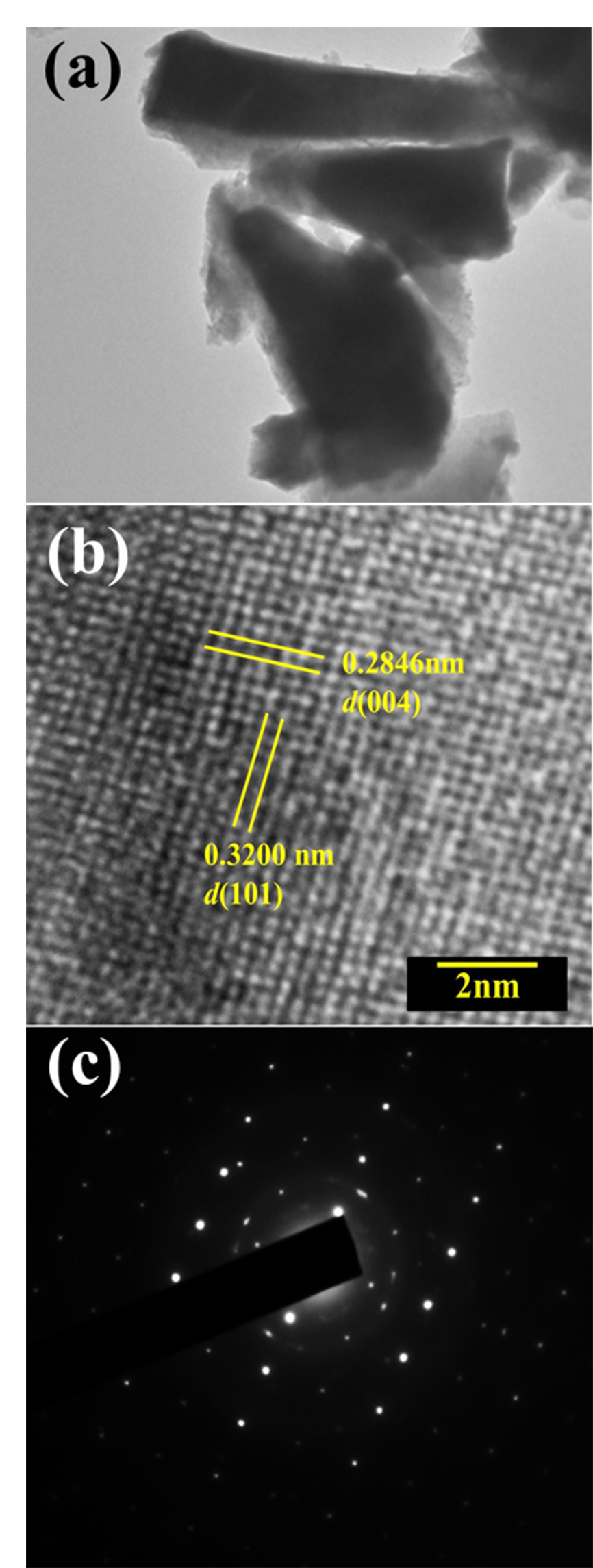}
    \caption{(a) TEM image of NbP single crystals is shown. (b) HRTEM image shows \textit{d}-spacings corresponding to (101) and (004) planes. The values of \textit{d}(101) and \textit{d}(004) match well with those calculated from XRD data. (c) SAED pattern of NbP confirm single crystalline growth.}
    
\end{minipage}
\end{figure}
Fig. 1 shows the refined diffraction pattern of powdered NbP crystals. The diffraction pattern is Rietveld refined using Fullprof software with both Nb and P occupying 4a Wyckoff positions (Nb (0, 0, 0) and P (0, 0, 0.4216)). The main panel of Fig. 1 shows the refined diffraction pattern of NbP. The sample crystallizes in a non-centrosymmetric space group I4$ _{1} $md.  The refined lattice parameters for NbP are found to be \textit{a} = 3.335\AA{} and \textit{c} = 11.379\AA{}, in agreement with reported results \cite{R20}. Inset (a) in Fig. 1 shows the optical image of NbP single crystalline sample in a millimetre scale. The sample was subjected to EDAX analysis to confirm elemental composition (see inset (b) of Fig. 1). The quantitative analysis by performing EDAX at several points provided the molar ratio Nb:P = 1:0.997 which confirms percentage composition of the sample close to its stoichiometric value. The schematic unit cell of the sample is also shown in the inset (c) of Fig. 1. Further, structural characterization of the sample was done using high resolution transmission electron microscope (HRTEM). Fig. 2 (a) shows the TEM image of the sample. The \textit{d}-values calculated from lattice fringes corresponding to planes (004) and (101) are 2.84 \AA{} and 3.28 \AA{}, respectively [Fig. 2(b)]. These values match well with the \textit{d}-spacing values obtained from XRD results.  Fig. 2 (c) shows the selected area diffraction pattern (SAED) for NbP that confirms excellent single crystalline characteristics of as grown samples.\par

Fig. 3 summarizes the temperature and field dependent magnetoresistivity measurements. Fig. 3 (a) shows temperature dependent resistivity measured in various transverse applied magnetic fields (\textit{H}$ \bot $\textit{I}) ranging from 0 to 6 T. In this configuration, magnetic field is applied along \textit{c}-axis. In zero applied field, the resistivity of NbP shows metallic behavior with \textit{$ \rho $}(300K) = 33 $ \mu\Omega $-cm and \textit{$ \rho $}(2.5 K) = 0.8 $ \mu\Omega $-cm. The residual resistivity ratio (RRR = \textit{$ \rho $}(300K)/\textit{$ \rho $}(2.5 K)) is estimated to be 41. Evidently, application of magnetic field induces extremely large magnetoresistance and changes the temperature dependence of resistivity significantly. An unambiguous broad shoulder in resistivity is also observed for fields above 1 T.  This upturn in resistivity exhibits semiconductor like behaviour before metallicity takes over at lower temperatures. This is indicative of opening of a gap in the presence of magnetic field that becomes more prominent at higher magnetic fields. Moreover, the crossover temperature shifts to higher temperature with increasing magnetic field. At low temperatures, saturation in electrical resistivity is observed even in the presence of magnetic field that demonstrates that the resistivity of the bulk material is short circuited by the dominance of metallic topological Fermi arcs at low temperatures \cite{R27, R28}. \par

The observed extremely large transverse MR can be better studied by isothermal measurement of field dependent resistivity. The magnetoresistance at temperature \textit{T} is calculated using the relation: MR = [(\textit{$\rho $}(\textit{H})-\textit{$ \rho $}(0)/\textit{$ \rho $}(0)] 100\%, where {$ \rho $}(\textit{H}) is the resistivity at field \textit{H} and {$ \rho $}(0) is the zero field resistivity. Figure 3(b) shows the field dependent MR at temperatures varying from 2.5 K to 100 K. Inset in Fig. 3(b) shows MR at 200 K and 300 K as well. Clear signatures of Shubnikov-de Haas quantum oscillations in electrical resistivity are observed. At 2.5 K, MR reaches 5.4$ \times $10$ ^{4}\%$ at 6 T without any trace of saturation. Even at room temperature large MR = 151\% is observed at 6 T. The high value of MR is comparable to recent reports of other Weyl semimetals \cite{R22, R26, R29}. In fact, MR shows weak temperature dependence upto 10 K and thereafter decreases sharply. Intriguingly, at low temperatures, the field dependent MR shows crossover from parabolic to linear dependence. At temperatures above 100 K, the field dependent MR shows parabolic behaviour. An interesting aspect of the data is the observation of almost linear magnetoresistance at high fields in both temperature regimes. \par 

\begin{figure*}
\centering
\begin{minipage}{1.0\textwidth}
\includegraphics[width=\textwidth]{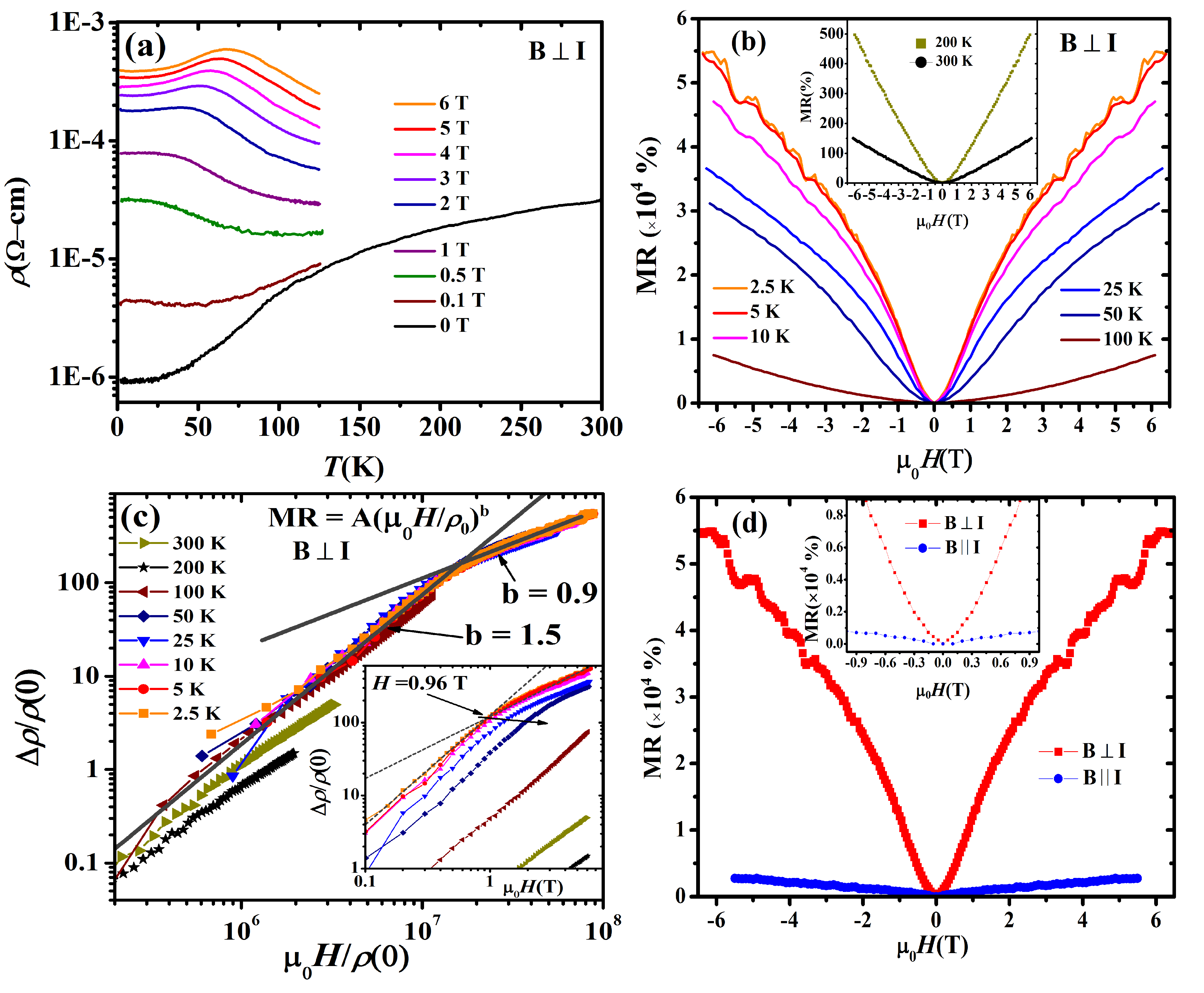}
    \caption{(a) Temperature dependent resistivity of NbP without field and in transverse applied magnetic (\textit{H}$ \bot $\textit{I}) fields (0 $ \textendash $ 6 T) is plotted as a function of temperature. Zero field resistivity shows metallic nature of the sample which shows an upturn on application of magnetic field followed by saturation. (b) Transverse isothermal magnetoresistance at different temperatures ranging from 2.5 K to 100 K. Shubnikov-de Haas (SdH) oscillations are clearly visible at 2.5 and 5 K. Inset in (b) shows the transverse magnetoresistance at 200 K and 300 K. (c) The log plot of the Kohler$'$s scaling of MR with reduced magnetic field ($ \mu_{0}\textit{H}/\rho(0)$). Inset shows the field dependent MR on log scale. (d) Shows the transverse and longitudinal magnetoresistance (\textit{H}$ \| $\textit{I}) are compared at 2.5 K. Inset in (d) The expanded view of low field region for clarity.}

\end{minipage}
\end{figure*}

Towards understanding the role of scattering in temperature dependent magnetoresistance, in Fig. 3 (c) we apply Kohler$'$s scaling to transverse MR data. According to Kohler$'$s rule \cite{R30}: $ \Delta \rho /\rho(0)= A(\textit{H}/\rho(0))^{b}$ , where $ \Delta \rho $ (=$ \rho $(\textit{H})-{$ \rho $}(0)) is the change in resistivity in transverse applied field and A and b are constants. In a semimetal, the charge carriers could show different relaxation time scales and effective masses with a spread in their velocities. Kohler$'$s rule predicts that when the charge carriers possess same scattering rate, the plot between MR vs \textit{H}/{$ \rho $}(0) curves measured at different temperatures merge into a single line. This is confirmed in Fig. 3(c) where for temperature up to 100 K, all the MR curves collapse into a single curve. This indicates that same scattering rate is followed by all the charge carriers in the sample. Moreover, at temperatures 200 K and 300 K, the curves do not fall on the same curve which can be associated with the change in concentration, sign and mobilities of compensated charge carriers. Further, a change in slope between high field and low field MR is observed for the collapsed curves upto 100 K. From the fit to Kohler$ ' $s relation in these two regimes, the values of parameter b is found to be 1.5 at low fields and 0.9 at high fields. This justifies a crossover from nearly parabolic behavior at low fields to linear behavior at high fields. From the plot between MR and \textit{H}, shown in the inset of Fig. 3(c), we observe that the crossover field increases on increasing temperature (~1 T at 2.5 K to ~2 T at 25 K). This suggests different scattering mechanisms at low and high magnetic fields are present at low temperatures and this feature has been seen in several other WSMs \cite{R31, R32}. The consequent changes in power law dependence of MR at different temperature are theoretically concluded in Weyl semimetals \cite{R33, R34}. \par 

The study of longitudinal MR in Weyl fermions is of substantial current interest as it brings forth the chirality aspects of electronic transport \cite{R21, R22, R35, R36, R37}.  Fig. 3(d) shows the comparative plots for both transverse and longitudinal field direction (with respect to applied current direction) at T = 2.5 K. At 5 T, MR(\textit{H}$ \bot $\textit{I} )/MR(\textit{H}$ \| $\textit{I}) =(4.75\texttimes10$ ^{4} $)/(0.27\texttimes10$^{4} $)= 17.3, which indicates that the MR in this material is relatively isotropic in comparison to layered semimetal WTe$ _{2} $ \cite{R38}. Further, in Weyl semimetals, negative longitudinal MR (LMR) is expected due to chiral anomaly effect \cite{R39,R40}. Clearly, no negative longitudinal MR is observed in NbP (see Fig. 3(d)). In recent theoretical studies, it is reported that there are two essential factors affecting the experimental observation of chiral anomaly effect in Weyl monopnictides \cite{R21, R27, R41}. First, it is required that the Weyl points lie close to the Fermi level and secondly, the chirality of the Weyl points needs to be well defined, i.e, each Weyl point of opposite chirality are enclosed by well separated Fermi pockets. Based on these facts, it is reported that in case of NbP and TaP, the experimental observation of chiral anomaly effect is unlikely \cite{R21, R41}. This is in contrast to LMR results in TaAs \cite{R21, R27, R41}. Negative LMR is also reported in NbP, TaP and NbAs \cite{R19, R42, R43}. However, the origin of the negative LMR is controversial as the geometry of voltage and current contacts on the sample can lead to current jetting effects which is not related to chiral anomaly \cite{R21, R37}. In case of NbP, it is known that there are two pairs of Weyl points, W1 and W2. W1 lies in $ k_{z} $ = 0 plane, ~ 57 meV below Fermi level while W2 lies in $ k_{z} $ = $ \pi/c $ plane ~ 5 meV above Fermi level. W1 and W2 are enclosed within the electron and hole pockets \cite{R41}. Since the Weyl point W1 is far away from the Fermi surface it may not contribute to chiral anomlay. Moreover, W2 is enclosed within the hole pocket and therefore it leads to the cancellation of chirality and Berry flux. Our null results on the absence of chiral anomaly therefore supports recent theoretical analysis on the improbability of experimental observation of chiral anomaly effect in NbP  \cite{R21, R27, R41}.\par

To gain more insight into charge transport in NbP, and to correlate the MR data with Fermi surface analysis, in Fig. 4(a) we show the field dependence of Hall resistance, R$ _{xy} $, at temperatures ranging from 2.5 K to 300 K. We observe that R$ _{xy} $ is negative in the temperature range 2.5 K to 100 K.  This indicates that the dominant quasiparticles on the Fermi surface are electron-like at low temperatures. At temperature above 100 K, R$ _{xy} $(\textit{H}) becomes positive, reflecting the nature of the quasiparticles has switched to hole-like behaviour with increasing temperature. A simple model to account for magneto-transport in semimetals would be to assume two types of carriers near the Fermi level. The electron-like quasiparticles come from the Weyl fermions that follow linear dispersion, with the Weyl cones placed slightly below the Fermi level. The hole-like quasiparticles on the other hand have a simple quadratic dispersion. Expanding the Hamiltonian near the Weyl cones in the low-energy limit, we can write the corresponding Hamiltonians as\par

\begin{equation}
\begin{multlined}
\textit{H}_{e}=-\nu(\sigma_{x}\textit{k}_{x}+\sigma_{y}\textit{k}_{y})+\textit{t}_{z} sin(\textit{k}_{z})\sigma_{z}\\
\approx-\nu(\sigma_{x}\textit{k}_{x}+\sigma_{y}\textit{k}_{y})+\nu_{z}\sigma_{z}\textit{k}_{z}
\end{multlined}
\end{equation}
\begin{equation}
\textit{H}_{h}=\frac{\hbar^{2}}{2m}(\textit{k}-\textit{k}_{0})^{2}
\end{equation}

Here $ \nu $ is the in-plane Fermi velocity, and $ \nu_{z} $ is the same but along the \textit{c}-axis. $ \textit{t}_{z} $ is the hoping integrals between the nearest neighbour lattice sites along the z-direction. Momentum \textbf{k} is measured with respect to the Weyl nodes. We notice that the Weyl Hamiltonian ($ \textit{H}_{e} $) is linear in momentum in all three directions with anisotropic Fermi velocities. The first term stems from the usual Weyl-like Hamiltonian as seen in other Weyl semimetals, while sin(\textit{k}$ _{z} $) ($ \sim $ \textit{k}$ _{z} $ in the low-energy region) is required to include due to the fact that Weyl nodes are present on both \textit{k}$ _{z} $ = 0 and \textit{k}$ _{z} $ = $ \pi/c $ planes. Therefore, unlike other Weyl/Dirac semimetals, in NbP one can expect chiral anomaly in all three directions.  However the Weyl points are far away from the Fermi level, so the unequivocal confirmation of chiral anomaly from magneto-transport in NbP is challenging.\par

The Hall data of Fig. 4(a) can be understood from the coexistence of relativistic electrons (linear dispersion) and non-relativistic holes (parabolic dispersion) with extremely high effective mass. Recalling that Hall resistivity is proportional to the density of states (DOS), in the following, we examine how the DOS qualitatively evolves with temperature for both Weyl electrons and heavy holes.  Evidently, the DOS of Weyl electrons is linear in energy, while that of the hole is quadratic, i.e.,  
\begin{equation}
\textit{N}_{e}(\textit{E})=b\textit{E}\approx b\textit{k}_{B}\textit{T}
\end{equation}
\begin{equation}
\textit{N}_{h}(\textit{E})=a\textit{E}^{2}\approx a\textit{k}_{B}\textit{T}^{2}
\end{equation}
here, a and b are constants. Since \textit{N}$_{e} $ is linear in \textit{T}, it dominates at low temperature regime. Above some threshold value of \textit{T}, the quadratic contributions from the hole part (\textit{N}$ _{h} $) becomes dominant. This behaviour is schematically shown in the inset of Fig. 4 (a). This qualitatively explains the sign reversal of the Hall resistivity data.
\par
Our attempt to fit a two band model to Hall data did not succeed and in the following we discuss the results of simplistic single band analysis. The Hall data for single-carrier Drude model is used to calculate charge carrier density ( \textit{N}$ _{e,h}$) and charge carrier mobility ($ \mu_{e,h} $ = R$ _{H} $(T)/$ \rho $(T)).  Here R$ _{H} $ is the Hall coefficient, \textit{N}$ _{e,h}$ is electron/hole charge carrier density, $ \mu_{e,h} $ is the average charge carrier mobility corresponding to electrons or holes, respectively, and $ \rho $(\textit{T}) is the resistivity of the sample. To calculate R$ _{H} $(\textit{T}), we have used slope of Hall resistance R$ _{xy} $(\textit{H}) at high fields. The value of R$ _{H} $, and \textit{N}$ _{e,h}$ thus calculated are shown in Fig. 4 (b). The value of electron charge carrier density at 2.5 K is found to be, \textit{N}$ _{e} $ = 3.5\texttimes10$ ^{18} $ cm$ ^{-3} $ which increases with increase in temperature demonstrating typical semimetallic behavior of the sample. The electron carrier mobility at 2.5 K is found to be very high, $ \mu_{e} $ = 2\texttimes10$ ^{6} $ cm$ ^{2} V ^{-1} s^{-1}$.  This value of electron charge carrier mobility is close to the reported results on NbP \cite{R19, R20}. \par

\begin{figure*}
\centering
\begin{minipage}{1.0\textwidth}
\includegraphics[width=\textwidth]{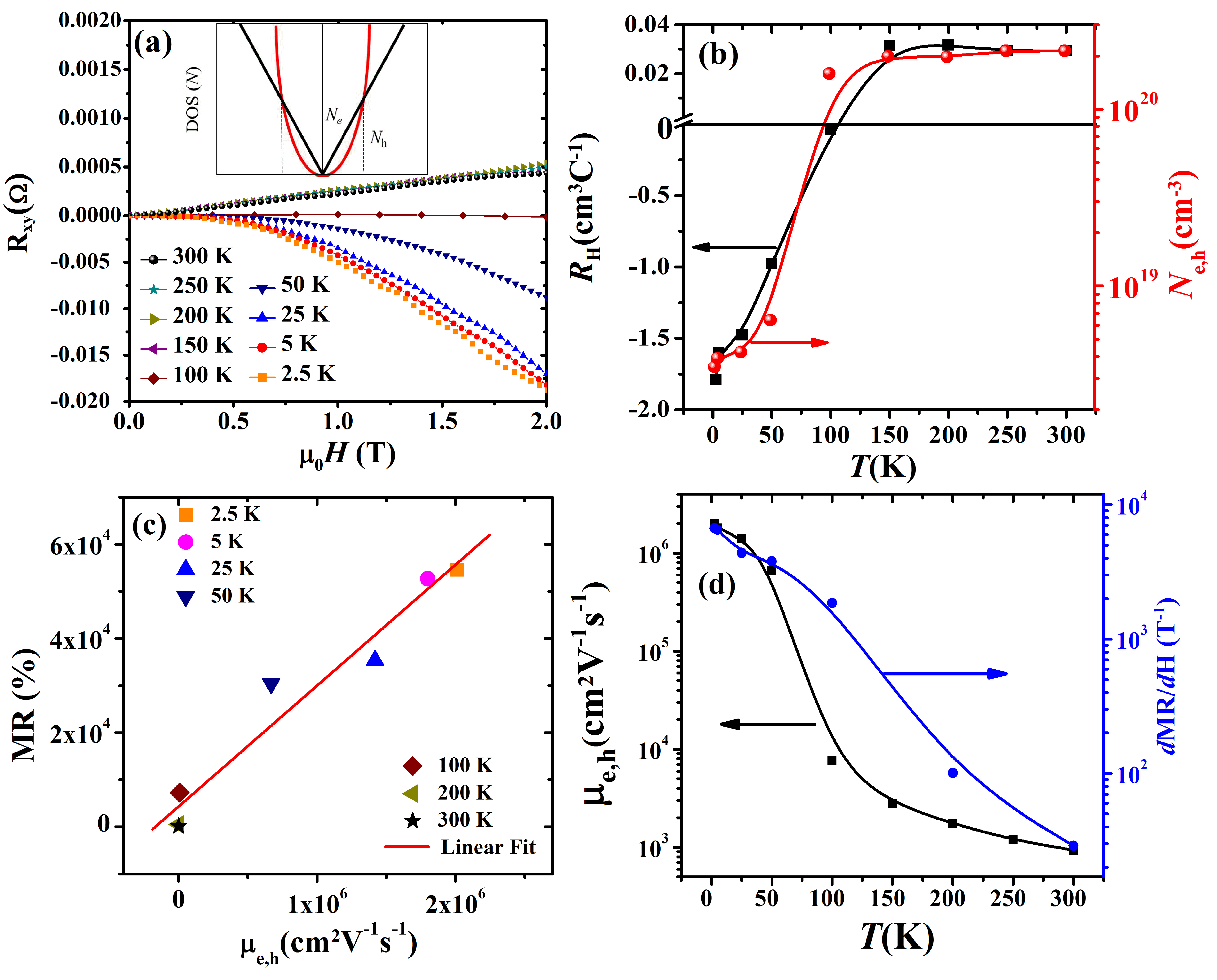}
     \caption{(a) Field dependent Hall resistance, R$ _{xy} $(\textit{H}) are plotted at different temperatures varying from 2.5 K to 300 K. We clearly see change of slope of R$ _{xy} $(\textit{H}) after 100 K. Inset shows the dependence of density of states (DOS) for electron and hole charge carriers on energy/temperature. It demonstrates that the dominant charge carriers are electrons at low temperatures and at high temperatures holes are the dominant charge carriers. (b) Hall coefficient (R$_{H} $) and charge carrier density ($ \textit{N}_{e,h} $) measured at temperatures ranging from 2.5 K to 300 K are shown. (c) Plot of MR (6 T) with respect to average mobility at varying temperatures. Straight line shows the linear fit to the data indicating linear dependence of MR variation on average mobility with temperature.  (d) Average mobility ($ \mu_{e,h} $)and \textit{d}MR/\textit{d}H (slope of MR at high fields) variation with temperature. The curves show same trend over temperature variation. The deviation near 100 K can be explained with the change of dominant charge carrier behavior in the sample.}
\end{minipage}
\end{figure*}

The observation of linear and non-saturating MR under high magnetic fields in WSMs is an interesting feature. Usually, MR varies as $ \textit{H}^{2} $ at low fields ($ < $ 1 T) due to Lorentz force deviation of charge carriers and shows saturation at high fields. The existence of linear non-saturating MR is highly intriguing. In metals, it may originate due to open Fermi surfaces (e.g. Au). In Weyl semimetals, however, this is not the case. Both quantum and classical origin of linear MR in metallic systems are proposed \cite{R44, R45, R46}. Quantum interpretation of linear non-saturating MR at high field was proposed by Abrikosov \cite{R44} at ultra-quantum limit. In this limit, only the first landau level (LL) is filled, which happens when $ \hbar\omega_{c} > E_{F} $ and $ E_{F} >> \kappa_{B}T $, where $ \hbar\omega_{c} $ = eB/m$ ^{\ast} $ is the cyclotron frequency. It gives the relation between carrier concentration and magnetic field as: n$  < $(eB/$  \hbar$)$ ^{3/2} $. From our Hall measurements, the value of carrier concentration n is $\sim 10^{18} cm^{-3} $ (T = 2.5 K) which gives the condition for ultra-quantum limit to be B$ > $7 T. However, our sample starts showing linear MR in field as low as 1 T which rules out the possibility of quantum origin of linear MR in our sample. The appearance of linear MR is therefore assigned to classical behaviour applicable to inhomogeneous systems with high charge carrier mobility. Parish and Littlewood (PL) \cite{R45, R46} proposed that linear MR originates from fluctuations in mobility in a strongly disordered system. According to PL model, MR strongly depends on fluctuations in mobility which is the 
ratio between width of mobility disorder $ \Delta\mu $ and average mobility  $\propto \langle\mu\rangle $. At high fields the PL equations for MR are as follows: MR $\propto \langle\mu\rangle $ for  $ \Delta\mu $/$ \langle\mu\rangle < 1$, and MR $\propto \Delta\mu $ for  $ \Delta\mu $/$ \langle\mu\rangle > 1$ .  As seen in Fig. 4 (c), the transverse MR varies linearly with average charge carrier mobility. This implies: MR $\propto \langle\mu\rangle $. Fig. 4(d) shows correlation between temperature dependent mobility and slope of MR in the high field region. From the plot, we observe that the change in MR follows the average mobility of charge carriers in the entire temperature range. This further confirms that MR $\propto \langle\mu\rangle $. The deviation between 100 K and 200 K are associated with change in type of charge carriers from electrons to holes. In summary, the linear MR in NbP is due to mobility fluctuations induced by scattering from low mobility inhomogeneous regions in the system.\par
As is strikingly evident in Fig. 3(b), in the presence of high transverse magnetic field, clear Shubnikov de-Haas (SdH) oscillations in MR are observed at low temperatures. The SdH oscillations are extracted by subtracting mean polynomial fitting to MR data. The oscillatory component ($ \Delta\rho $) thus extracted is shown in Fig. 5 (inset (a)) as a function of $ 1/\mu_{0}\textit{H} $. The oscillations show periodicity with $ 1/\mu_{0}\textit{H} $. The spectrum is substantially complex due to the contribution from various sub-bands on the Fermi surface. The SdH oscillations are further analysed using the following expression for the oscillatory component of magnetoresistance in a 3D system:

\begin{equation}
\Delta\rho/\langle\rho(0)\rangle =A(\textit{T},\mu_{0}\textit{H})\cos[2\pi(\frac{F}{\mu_{0}\textit{H}} -\gamma+\delta)]
\end{equation}

where $ \langle\rho(0)\rangle $ is the non-oscillatory component of transverse $ \rho $, F is the frequency of oscillation, $ \gamma $ is the Onsager phase and $ \delta $ (= $ \pm $1/8) is the phase shift introduced by the three-dimensionality of the Fermi-surface. By performing Fast-Fourier transformation (FFT) of $ \Delta\rho $ vs 1/$ \mu_{0}\textit{H} $ data, we have extracted three oscillations corresponding to frequencies: $ F_{\alpha} $ = 7 T, $ F_{\beta} $ = 13.6 T and $ F_{\gamma} $ = 30.5 T with its harmonic at 61 T, (Fig. 5 (b)). Frequencies $ F_{\alpha}$ , $ F_{\beta}$  and $ F_{\gamma} $  are associated with the electron pockets near the Fermi energy in NbP \cite{R41}. In the present case, the period of oscillations corresponding to $ F_{\gamma} $ (1/$ \mu_{0}\textit{H} $) is 0.033 T$ ^{-1} $.  Further, the frequency of quantum oscillations is proportional to the cross-sectional Fermi-surface area, A$ _{F} $, that follows the Lifshitz-Onsager relation: $ F = (\Phi_{0}/(2\pi^{2}))A_{F} $ . Here, $ \Phi_{0} $ = h/2e is the magnetic flux quantum and e is the electronic charge. The calculated cross-sectional area of the Fermi-surfaces corresponding to frequencies $ F_{\alpha}$, $ F_{\beta}$ and $ F_{\gamma} $ are 0.21\texttimes10$ ^{-3} \AA{}^{2}$, 0.41\texttimes10$ ^{-3} \AA{}^{2}$ and 3\texttimes10$ ^{-3} \AA{}^{2}$, respectively. The cross-sectional Fermi-surface for $ F_{\gamma} $ matches well with the reported data \cite{R20}. These areas are very small, only about 0.006\%, 0.01\% and 0.08\% of the total area of the Brillouin zone in the $ k_{x} $-$ k_{y} $ plane, respectively. The Fermi wave-vectors are then calculated from the observed $ A_{F} $ values using the relation: $ k_{F} = (A_{F}/\pi)^{1/2} $. The calculated $ k_{F} $ corresponding to $ F_{\alpha}$, $ F_{\beta}$ and $ F_{\gamma} $ are 8\texttimes10$ ^{-3} \AA{}^{-1}$, 0.011 $\AA{}^{-1}$ and 0.031 $\AA{}^{-1}$, respectively. \par 

\begin{figure}
\centering
    \includegraphics[width=0.45\textwidth]{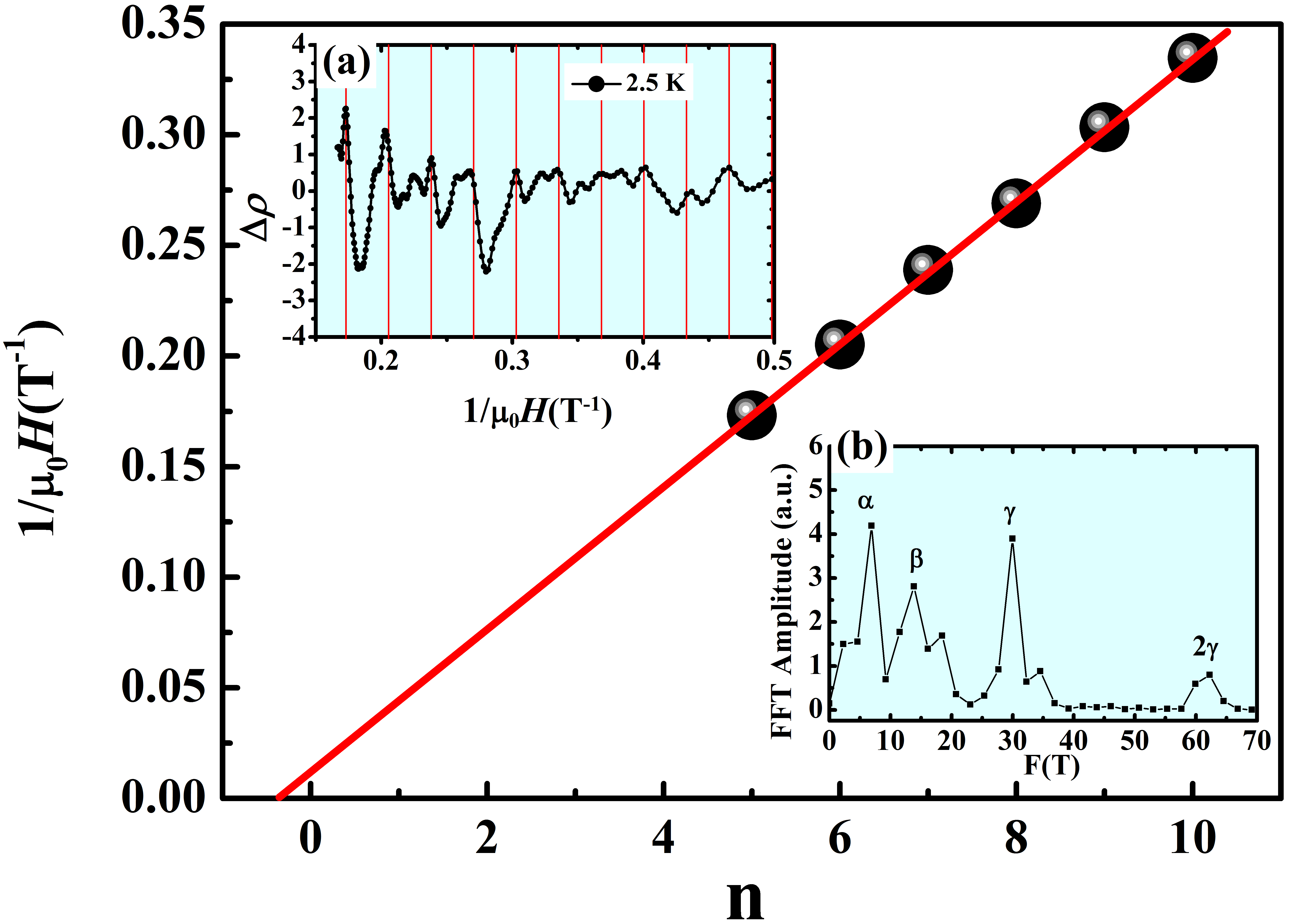}
    \caption{Inset (i) shows the Shubnikov-de Haas (SdH) oscillations extracted by subtracting the background from MR measured at 2.5 K and 5 K. Background is calculated by fitting the data with a polynomial function. Gray dot-dashed grid lines are placed to see the period of oscillation clearly. The period shown in figure is 0.033 T$^{-1} $ corresponding to F$ \gamma $ frequency of SdH oscillations. Inset (ii) shows the Fast Fourier transform (FFT) amplitude of SdH oscillations measured at 2.5 K. Main figure shows the Landau fan diagram corresponding to F$ \gamma $ frequency of SdH oscillations. The data is fit linearly and it gives the value of intercept 0.31 which indicates trivial Berry phase for F$ \gamma $.}
\end{figure}
From the quantum oscillation data, in the following we discuss the topological signatures of these electron and hole pockets by studying the Berry phase accumulated around the Weyl nodes. Identifying the Berry phase ($ \Phi_{B} $) corresponding to Fermi pockets in Weyl semimetal is challenging because of its multi band characteristics.  The Onsager phase $ \gamma=1/2-\Phi_{B}/2\pi+\delta $, relates to $ \Phi_{B} $.  In case of topologically trivial band, $ \Phi_{B} $ = 0, which implies $ \gamma= 1/2+\delta $ while in case of topologically non-trivial band, $ \Phi_{B} = \pi $, which implies $ \gamma = \delta $. To identify the topological nature of the Fermi-surface associated with the most prominent frequency in our data, $ F_{\gamma} $, we have plotted the Landau fan diagram (a plot between Landau Level (LL) index n and $ 1/\mu_{0}\textit{H} $) in main panel of Fig. 5 (T = 2.5 K). The linear fitting gives Onsager phase of $ \gamma $ = 0.31. For electron pockets, $ \delta $ = -1/8 and therefore the Berry phase parameter, $ \Phi_{B} \sim 0$ indicating a topologically trivial Berry phase in NbP corresponding to frequency $ F_{\gamma} $.  This is in disagreement with reported conclusion on non-trivial Berry phase in NbP \cite{R19}.\par
\section{\label{sec:level2}Conclusions}
In conclusion, we provide evidence for extremely high, non-saturating linear transverse magnetoresistance in high quality single crystals of Weyl semimetal NbP.  Clear Shubnikov de Haas quantum oscillations are observed at high fields and low temperatures with three main frequencies. The Berry phase calculated for frequency = 30.5 T of SdH oscillations shows null additional phase factor. No evidence for longitudinal magneto-resistance is seen that rules out chiral anomaly in NbP.  This is ascribed to large separation between Fermi level and Weyl nodes and lack of well defined chirality.  Combined Hall and magneto-transport data suggest that the linear MR observed in NbP is due to charge carrier mobility fluctuations.

\section{\label{sec:level2}Acknowledgements}
Sudesh, P. Kumar and P. Neha acknowledge UGC-D S Kothari fellowship, JNU, New Delhi and UGC-BSR, respectively for financial support. Authors are thankful to AIRF (JNU) for access to the PPMS, EDAX and TEM facilities. We thank Dr. Chandrashekhar for useful discussions. Low – temperature high magnetic field at JNU is supported under the FIST program of Department of Science and Technology, Government of India. TD’s work is supported by SERB young scientist Startup research grant.


\begin{thebibliography}{50}

\bibitem{R1}X. Wan, A. M. Turner, A. Vishwanath, and S. Y. Savrasov, Phys. Rev. B \textbf{83}, 205101 (2011).
\bibitem{R2}B. Singh, A. Sharma, H. Lin, M. Z. Hasan, R. Prasad, and A. Bansil, Phys. Rev. B \textbf{86}, 115208 (2012).
\bibitem{R3}A. M. Turner and A. Vishwanath, arXiv:1301.0330v1 (2013).
\bibitem{R4}Y. Xia, D. Qian, D. Hsieh, L. Wray, A. Pal, H. Lin, A. Bansil, D. Grauer, Y. S. Hor, R. J. Cava, and M. Z. Hasan, Nat. Phys.\textbf{5}, 398 (2009).
\bibitem{R5}H. Zhang, C.-X. Liu, X.-L. Qi, X. Dai, Z. Fang, and S.-C. Zhang, Nat. Phys. \textbf{5}, 438 (2009).
\bibitem{R6}Y. L. Chen, J. G. Analytis, J.-H. Chu, Z. K. Liu, S.-K. Mo, X.-L. Qi, H. J. Zhang, D. H. Lu, X. Dai, Z. Fang, S. C. Zhang, I. R. Fisher, Z. Hussain, and Z.-X. Shen, Science \textbf{325}, 178 (2009).
\bibitem{R7}Q. D. Gibson, L. M. Schoop, L. Muechler, L. S. Xie, M. Hirschberger, N. P. Ong, R. Car, and R. J. Cava, Phys. Rev. B \textbf{91}, 205128 (2015).
\bibitem{R8}X. Wan, A. M. Turner, A. Vishwanath, and S. Y. Savrasov, Phys. Rev. B \textbf{83}, 205101 (2011).
\bibitem{R9}K. S. Novoselov, A. K. Geim, S. V Morozov, D. Jiang, M. I. Katsnelson, I. V Grigorieva, S. V Dubonos, and A. A. Firsov, Nature \textbf{438}, 197 (2005).
\bibitem{R10}H. Weng, C. Fang, Z. Fang, B. A. Bernevig, and X. Dai, Phys. Rev. X \textbf{5}, 011029 (2015).
\bibitem{R11}L. X. Yang, Z. K. Liu, Y. Sun, H. Peng, H. F. Yang, T. Zhang, B. Zhou, Y. Zhang, Y. F. Guo, M. Rahn, D. Prabhakaran, Z. Hussain, S.-K. Mo, C. Felser, B. Yan, and Y. L. Chen, Nat. Phys. \textbf{11}, 728 (2015).
\bibitem{R12}Z. K. Liu, L. X. Yang, Y. Sun, T. Zhang, H. Peng, H. F. Yang, C. Chen, Y. Zhang, Y. F. Guo, D. Prabhakaran, M. Schmidt, Z. Hussain, S.-K. Mo, C. Felser, B. Yan, and Y. L. Chen, Nat. Mater. \textbf{15}, 27 (2016).
\bibitem{R13}K. Y. Yang, Y. M. Lu, and Y. Ran, Phys. Rev. B \textbf{84}, 075129 (2011).
\bibitem{R14}A. A. Burkov and L. Balents, Phys. Rev. Lett. \textbf{107}, 127205 (2011).
\bibitem{R15}K. Fukushima, D. E. Kharzeev, and H. J. Warringa, Phys. Rev. D \textbf{78}, 074033 (2008).
\bibitem{R16}D. T. Son and B. Z. Spivak, Phys. Rev. B \textbf{88}, 104412 (2013).
\bibitem{R17}A. A. Taskin and Y. Ando, Phys. Rev. B \textbf{84}, 035301 (2011).
\bibitem{R18}J. Hu, J. Y. Liu, D. Graf, S. M. A. Radmanesh, D. J. Adams, A. Chuang, Y. Wang, I. Chiorescu, J. Wei, L. Spinu, and Z. Q. Mao, Sci. Rep. \textbf{6}, 18674 (2016).
\bibitem{R19}Z. Wang, Y. Zheng, Z. Shen, Y. Lu, H. Fang, F. Sheng, Y. Zhou, and X. Yang, Phys. Rev. B \textbf{93}, 121112(R) (2016).
\bibitem{R20}C. Shekhar, A. K. Nayak, Y. Sun, M. Schmidt, M. Nicklas, I. Leermakers, U. Zeitler, Y. Skourski, J. Wosnitza, Z. Liu, Y. Chen, W. Schnelle, H. Borrmann, Y. Grin, C. Felser, and B. Yan, Nat. Phys. \textbf{11}, 645 (2015).
\bibitem{R21}F. Arnold, C. Shekhar, S. Wu, Y. Sun, R. D. dos Reis, N. Kumar, M. Naumann, M. O. Ajeesh, M. Schmidt, A. G. Grushin, J. H. Bardarson, M. Baenitz, D. Sokolov, H. Borrmann, M. Nicklas, C. Felser, E. Hassinger, and B. Yan, Nat. Commun. \textbf{7}, 11615 (2016).
\bibitem{R22}L. P. He, X. C. Hong, J. K. Dong, J. Pan, Z. Zhang, J. Zhang, and S. Y. Li, Phys. Rev. Lett. \textbf{113}, 246402 (2014).
\bibitem{R23}Z. Wang, Y. Zheng, Z. Shen, Y. Zhou, X. Yang, Y. Li, C. Feng, and Z.-A. Xu, Phys. Rev. B \textbf{93}, 121112(R) (2016).
\bibitem{R24}S.-M. Huang, S.-Y. Xu, I. Belopolski, C.-C. Lee, G. Chang, B. Wang, N. Alidoust, G. Bian, M. Neupane, C. Zhang, S. Jia, A. Bansil, H. Lin, and M. Z. Hasan, Nat. Commun. \textbf{6}, 7373 (2015).
\bibitem{R25}D. Kumar and A. Lakhani, arXiv:1606.09059v1 (2016).
\bibitem{R26}C. Shekhar, V. Sü, and M. Schmidt, arXiv:1606.06649 (2016).
\bibitem{R27}F. Arnold, M. Naumann, S.-C. Wu, Y. Sun, M. Schmidt, H. Borrmann, C. Felser, B. Yan, and E. Hassinger, arXiv:1603.08846v1 (2016).
\bibitem{R28}G. Chang, S. Xu, H. Zheng, C. Lee, S. Huang, I. Belopolski, D. S. Sanchez, G. Bian, N. Alidoust, T. Chang, C. Hsu, H. Jeng, A. Bansil, H. Lin, and M. Z. Hasan, Phys. Rev. Lett. \textbf{116}, 066601 (2016).
\bibitem{R29}N. J. Ghimire, A. S. Botana, D. Phelan, H. Zheng, and J. F. Mitchell, arXiv:1604.04232v1 (2016).
\bibitem{R30}J. M. Ziman, \textit{Principles of the Theory of Solids.} (Cambridge university press, 1972).
\bibitem{R31}C. Zhang, C. Guo, H. Lu, X. Zhang, Z. Yuan, Z. Lin, J. Wang, and S. Jia, Phys. Rev. B \textbf{92}, 041203 (2015).
\bibitem{R32}C. Zhang, Z. Yuan, S. Xu, Z. Lin, B. Tong, M. Z. Hasan, J. Wang, C. Zhang, and S. Jia, arXiv:1502.00251 (2015).
\bibitem{R33}K. Ziegler, arXiv:1501.00268v1 (2015).
\bibitem{R34}N. Ramakrishnan, M. Milletari, and S. Adam, Phys. Rev. B \textbf{92}, 245120 (2015).
\bibitem{R35}X. Huang, L. Zhao, Y. Long, P. Wang, D. Chen, Z. Yang, H. Liang, M. Xue, H. Weng, Z. Fang, X. Dai, and G. Chen, Phys. Rev. X \textbf{5}, 031023 (2015).
\bibitem{R36}C.-Z. Li, L.-X. Wang, H. Liu, J. Wang, Z.-M. Liao, and D.-P. Yu, Nat. Commun. \textbf{6}, 10137 (2015).
\bibitem{R37}R. D. dos Reis, M. O. Ajeesh, N. Kumar, F. Arnold, C. Shekhar, M. Naumann, M. Schmidt, M. Nicklas, and E. Hassinger, New J. Phys. \textbf{18}, 085006 (2016).
\bibitem{R38}Y. Zhao, H. Liu, J. Yan, W. An, J. Liu, X. Zhang, H. Wang, Y. Liu, H. Jiang, Q. Li, Y. Wang, X. Z. Li, D. Mandrus, X. C. Xie, M. Pan, and J. Wang, Phys. Rev. B \textbf{92}, 041104(R) (2015).
\bibitem{R39}S. L. Adler, Phys. Rev. \textbf{177}, 2426 (1969).
\bibitem{R40}J. S. Bell and R. Jackiw, Nuov Cim A \textbf{60}, 47 (1969).
\bibitem{R41}J. Klotz, S. C. Wu, C. Shekhar, Y. Sun, M. Schmidt, M. Nicklas, M. Baenitz, M. Uhlarz, J. Wosnitza, C. Felser, and B. Yan, Phys. Rev. B \textbf{93}, 121105(R) (2016).
\bibitem{R42}J. Du, H. Wang, Q. Mao, R. Khan, B. Xu, Y. Zhou, Y. Zhang, J. Yang, B. Chen, C. Feng, and M. Fang, Sci. China Phys. Mech. Astron. \textbf{59}, 657406 (2016).
\bibitem{R43}X. Yang, Y. Liu, Z. Wang, Y. Zheng, and Z. Xu, arXiv:1506.03190v1 (2015).
\bibitem{R44}A. A. Abrikosov, Phys. Rev. B \textbf{58}, 2788 (1998).
\bibitem{R45}M. M. Parish and P. B. Littlewood, Nature \textbf{426}, 162 (2003).
\bibitem{R46}M. M. Parish and P. B. Littlewood, Phys. Rev. B \textbf{72}, 094417 (2005).




\end{thebibliography}
\end{document}